\documentclass[twocolumn]{revtex4}%
\usepackage{amsfonts}
\usepackage{amsmath}
\usepackage{amssymb}
\usepackage{graphicx}
\begin{document}
\title[Scarring in a driven system]{Scarring in a driven system with wave chaos}
\author{A.L. Virovlyansky}
\affiliation{Institute of Applied Physics, 46 Ulyanov St., Nizhny Novgorod 603950, Russia }
\author{G.M. Zaslavsky }
\affiliation{Courant Institute of Mathematical Sciences, New York University 251 Mercer
St., New York, NY 10012 and Department of Physics, New York University 2-4
Washington Place, New York, NY 10003, USA }

\begin{abstract}
We consider acoustic wave propagation in a model of a deep ocean acoustic
waveguide with a periodic range-dependence. Formally, the wave field is
described by the Schr\"{o}dinger equation with a time-dependent Hamiltonian.
Using methods borrowed from the quantum chaos theory it is shown that in the
driven system under consideration there exists a \textquotedblleft
scarring\textquotedblright\ effect similar to that observed in autonomous
quantum systems.

\end{abstract}
\maketitle

\bigskip It is well-known in the theory of quantum chaos that classical
periodic orbits, to a significant extent, determine both the
distribution of energy levels and structures of eigenfunctions
\cite{Gutz90,Reichl92,Stock99}. In particular, for some
eigenfunctions, the amplitudes are high in the vicinity of
unstable periodic orbits and low elsewhere. This phenomenon, first
discovered in a quantum billiard, is called the scarring
\cite{Heller84}. It has been established that the scarring is a
common property for many autonomous chaotic systems
\cite{Stock99,BKH2001,TP2004}. The description of scarring is
related to the problem of construction of eigenfunctions on the
basis of purely semiclassical calculations
\cite{Bogomolny88,Berry88,Berry92,Agam94,TH93a,TH93b}.

It is also well-known that so different physical problems as
quantum particle dynamics and wave propagation in nonuniform media
in the narrow beam approximation are described by the same
parabolic (or Shr\"{o}dinger) equation. One of the most intriguing
object is the underwater acoustic (UWA) and the long range sound
propagation in the ocean \cite{AET1,V2003b,V2003c}. Application of
methods of quantum chaos to the wave chaos proves to be efficient
in many publications \cite{AZ81,AZ91,Abdullaev,V2000a,SZ99}. The
goal of this paper is to extend this analogy between quantum and
wave chaos and to present simulations that demonstrate the
phenomenon of scarring in a nonautonomous Hamiltonian system with
1.5 degrees of freedom related to the UWA. We consider a
simplified case of wave propagation in a two dimensional
range-dependent model of hydroacoustic waveguide that has been
used to study wave and ray chaos at long range sound propagation
in the ocean \cite{SBT92a,V2001,V2002,V2004a,ZA97}. Formally, the
simplified model of ray propagation is equivalent to a nonlinear
oscillator perturbed by a periodic force, while the corresponding
parabolic equation is equivalent to the Shr\"{o}dinger equation
for the oscillator.

Consider a two-dimensional UWA waveguide with the sound speed $c$ as a
function of depth, $z$, and range, $r$. In the parabolic equation
approximation (valid under assumptions that waves propagate at small grazing
angles) the monochromatic wave field $u$ at a carrier frequency $f$ obeys the
parabolic equation \cite{Tappert77,BL91}%

\begin{equation}
\frac{i}{k}\frac{\partial u}{\partial r}=\hat{H}u,\;\hat{H}=-\frac{1}{2k^{2}%
}\frac{\partial^{2}}{\partial z^{2}}+U(r,z), \label{parabolic}%
\end{equation}
where $U(r,z)=\left(  1-\bar{c}^{2}/c^{2}(r,z)\right)/2.$ Here
$\bar{c}$ and $k=2\pi f/\bar{c}$ are a reference sound speed and a
reference wavenumber, respectively. Equation (\ref{parabolic})
formally coincides with the time-dependent Schr\"{o}dinger
equation. In this analogy $r$ plays a role of time, $\hat{H}$ is
the Hamiltonian operator, $U(r,z)$ is the potential, and $k^{-1}$
associates with the Planck constant. The ray paths are determined
by the standard Hamilton equations $dp/dt=-\partial H/\partial z$
and $dz/dr=\partial H/\partial p$ with the Hamiltonian
$H=p^{2}/2+U(r,z)$, where the momentum $p$ is related to the ray
grazing angle $\theta$ by $p=\tan\theta$.

Following Ref. \cite{SBT92a} we take the sound speed field $ c(r,z)=c_{0}%
(z)+\delta c(r,z)$, where $c_{0}(z)$ is the so-called Munk profile widely used
to model sound propagation in the deep sea \cite{BL91},
\begin{equation}
c_{0}(z)=\bar{c}\left(  1+\varepsilon(\left(  e^{-2(z-z_{a})/B}+2(z-z_{a}%
)/B-1\right)  \right)  \label{Munk}%
\end{equation}
with $\bar{c}=1.5$ km/s, $z_{a}=-1$ km, $\varepsilon=0.0057$, and $B=1$ km,
and
\begin{equation}
\delta c(r,z)=-2\gamma\,\bar{c}\,z/B\,e^{-2z/B}\,\sin(2\pi r/L).
\label{deltacM}%
\end{equation}
with $\gamma=0.01$ and $L=10$ km is a perturbation that generates
the ray chaos. The corresponding potential $U(r,z)$ is a periodic
function of range. Its shapes at three different ranges are shown
in Fig. 1. \begin{figure}[h]
\begin{center}
\includegraphics[
height=4.8193cm, width=7.cm ]{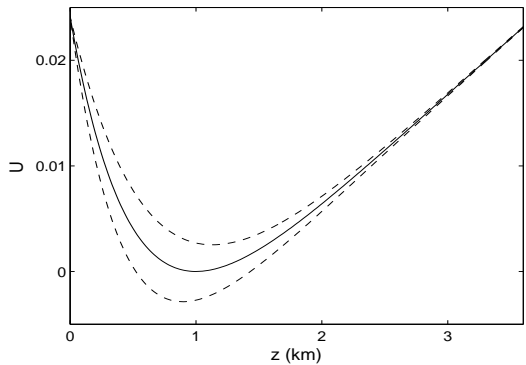}
\end{center}
\caption{Unperturbed potential $U(0,z)$ (solid line) and its maximal
deviations (dashed line) due to the perturbation.}%
\end{figure}

Figure 2 presents the phase portrait of the ray system plotted
using the Poincar\'{e} map
\cite{AZ91,PBTB88,Abdullaev,BTG91,SBT92a} $
(p_{n+1},z_{n+1})=\hat{T}(p_{n},z_{n})$, where $\hat{T}$ denotes
transformation of the momentum and coordinate of a ray trajectory
taken at range $nL$ to range $(n+1)L$ for $n=0,1,\ldots$. We see a
typical picture of stable islands formed by regular curves
submerged in a chaotic sea filled with randomly scattered points
depicting chaotic rays. \begin{figure}[hptbh]
\begin{center}
\includegraphics[
height=8.1015cm, width=5.7047cm,angle=-90 ]{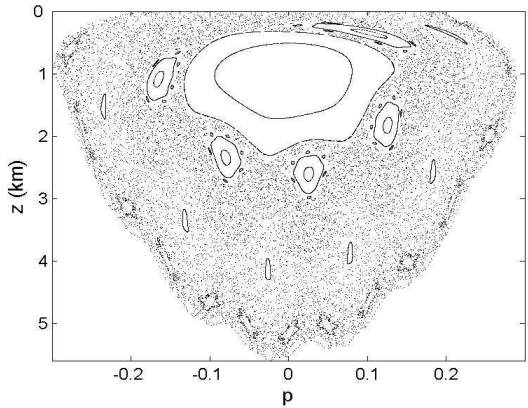}
\end{center}
\caption{Phase portrait of the ray system.}%
\end{figure}

To find out how the structure in Fig. 2 reveals itself in a wave dynamics
described by Eq. (\ref{parabolic}) we shall use the Floquet theorem
\cite{Reichl92,Stock99} and follow Ref. \cite{GPF84} where a similar approach
was applied. The solution of Eq. (\ref{parabolic}%
) can be presented as a sum of Floquet modes
\begin{equation}
u_{m}(r,z)=e^{i\alpha_{m}r/L}\Psi_{m}(r,z), \label{u-m}%
\end{equation}
where $\Psi_{m}(r+L,z)=\Psi_{m}(r,z)$ are range-periodic, $\alpha_{m}$ are
real constants, and $m=0,1,\ldots$. In quantum mechanics analogous terms
describe the so-called Floquet states, and constants $\alpha_{m}$ are
proportional to the quasi-energies. $\Psi_{m}(r,z)~$\ are eigenfunctions of
the shift operator $\hat{F}$: $\hat{F}\Psi_{m}(r,z)=\Psi_{m}(r+L,z)=e^{i\alpha
_{m}}\Psi_{m}(r,z)$. Since $\hat{F}$ is Hermitian, $\Psi_{m}(r,z)$ form a
complete set \cite{Reichl92,Stock99}.

Let us consider functions $\psi_{m}(z)\equiv\Psi_{m}(0,z)$. They
can be presented as a decomposition $
\psi_{m}(z)=\sum_{q}C_{mq}\varphi_{q}(z)$, where $\varphi_{q}(z)$
($q=0,1,\ldots$) are solutions to Sturm-Liouville eigenvalue
problem $\hat{H}_{0}\varphi_{m}=E_{m}\varphi_{m}(z)$ in the
unperturbed waveguide with the Hamiltonian $\hat{H}_{0}$
corresponding to $c(r,z)=$ $c_{0}(z)$. The eigenfunctions
$\varphi_{m}(z)$ and eigenvalues $E_{m}$ are readily found using a
standard mode code \cite{JKPS94}. It is easy to see that the phase
factors $e^{i\alpha_{m}}$ and columns of matrix $\left\Vert
C_{mq}\right\Vert $ are eigenvalues and eigenvectors,
respectively, of a unitary matrix with
elements%
\begin{equation}
F_{mq}=\int dz~\varphi_{m}(z)\hat{F}\varphi_{q}(z). \label{Fmn}%
\end{equation}
Note that $\hat{F}\varphi_{q}(z)$ is a solution to the parabolic equation
(\ref{parabolic}) at range $r=L$ found for an initial condition
$u(0,z)=\varphi_{q}(z)$. We have solved Eq. (\ref{parabolic}) using the code
MMPE \cite{MMPE}. Numerical results presented below have been obtained for a
carrier frequency $f=200$ Hz.

Associating a parameter $\mu$ with each function $\psi_{m}(z)$,
where $\mu=\sum_{q}q\left\vert C_{mq}\right\vert ^{2}$ is sum of
intensity weighted numbers of eigenfunctions $\varphi_{q}$, we
sort the Floquet modes in the order of increasing values of $\mu$.
Therefore the eigenfunctions $\psi_{m}(z)$ with small $m$ describe
waves localized at depths $z$ in the vicinity of the minimum of
$U(0,z)$. The greater is the mode number $m$ the steeper are the
grazing angles of waves forming the Floquet mode.

In order to link the structure of the Floquet mode with the classical phase
space we shall consider the Husimi distribution function \cite{SZ99,Reichl92}
\[
w_{m}(p,z)=\left\vert \frac{1}{\sqrt[4]{2\pi\Delta_{z}^{2}}}\int dz^{\prime
}\psi_{m}(z^{\prime})\right.
\]%
\begin{equation}
\times\left.  \exp\left[  ikp(z^{\prime}-z)-\frac{(z^{\prime}-z)^{2}}%
{4\Delta_{z}^{2}}\right]  ^{2}\right\vert \label{projection}%
\end{equation}
representing a projection of the wave field onto a Gaussian state
with minimum uncertainty (coherent state) centered at a point
$(p,z)$. Since the ray grazing angle $\theta$ with respect to the
$r$-axis is related to the momentum by $\theta=\arctan p$ the
distribution $w_{m}(p,z)$ may be interpreted as a local spectrum
of the wave field smoothed over spatial and angular
scales.\begin{figure}[h]
\begin{center}
\includegraphics[
trim=0.5in 0.in 0.in 0.in, height=8.5254cm, width=8.5cm ]
{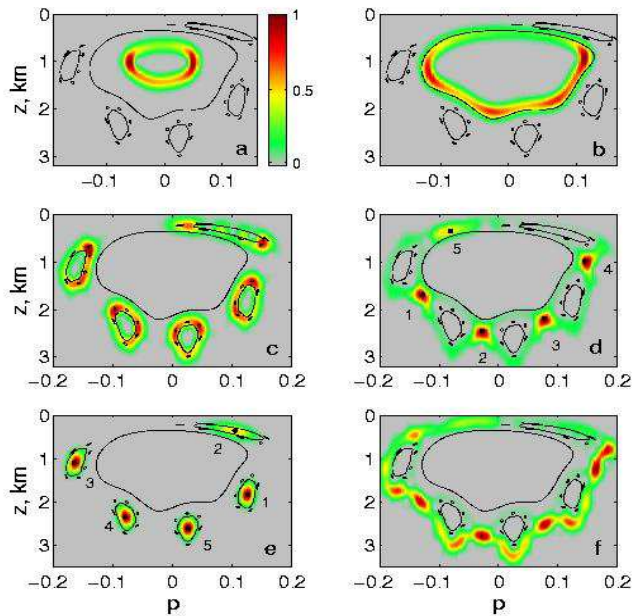}
\end{center}
\caption{Husimi distributions for the Floquet eigenfunctions $\psi_{10}(z)$
(a), $\psi_{40}(z)$ (b), $\psi_{66}(z)$ (c), $\psi_{77}(z)$ (d), $\psi
_{88}(z)$ (e), and $\psi_{105}(z)$ (f). Parameter $\Delta_{z}=0.01$ km.}%
\end{figure}

Figure 3 shows Husimi functions evaluated for six Floquet modes.
These examples have been selected to demonstrate a variety of
structures of Floquet modes in our environmental model. At each
panel we have outlined borders of stable islands discernible in
the central part of the phase portrait. Although the range
variations of the potential $U$ are most pronounced at its
minimum, the first Floquet eigenfunctions $\psi_{m}(z)$ are most
closed to their unperturbed counterparts, i.e. to functions
$\varphi_{m}(z)$ with the same numbers $m$. In the WKB
approximation the Husimi function for any $\varphi_{m}(z)$ is a
fuzzy version of a closed curve representing a period of an
unperturbed ray whose Hamiltonian is equal to $E_{m}$. In Fig. 3a
we see an example of such a fuzzy curve. Here a Husimi function
for $\psi_{10}(z)$ is plotted. It almost coincides with the
unperturbed Husimi function for $\varphi_{10}(z)$ (not shown). The
Husimi functions for the first 50 eigenfunctions $\psi_{m}(z)$ are
localized within the central island. The fuzzy curves
corresponding to $m$ approaching $50$ repeat the shape of the
border of the central island. This is clearly seen in Fig. 3b
where the Husimi function constructed for $\psi _{40}(z)$ is
shown.\begin{figure}[h]
\begin{center}
\includegraphics[
height=5.297cm, width=7.2cm ]{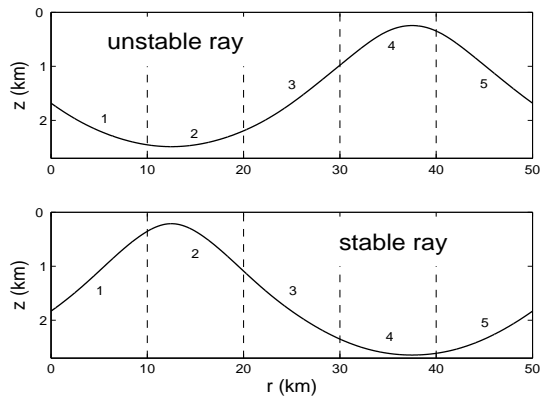}
\end{center}
\caption{Unstable (upper panel) and stable (lower panel) periodic
rays with
the period 50 km.}%
\end{figure}

Figures 3b-f demonstrate that the Husimi functions may have quite
different structures. But there exists a remarkable common feature
typical for all the first 140 Floquet modes considered in our
numerical simulations: each distribution is localized within (i)
the central stable island (Figs. 3a and b), (ii) within five
sub-islands surrounding the central one (Fig. 3e), or (iii) within
the chaotic sea (Figs. 3c, d, and f). Thus, the structure of
Husimi functions correlates with the shape of stable islands.
Roughly speaking, one may say that there are \textquotedblleft
chaotic\textquotedblright\ and \textquotedblleft
regular\textquotedblright \ Floquet modes whose Husimi functions
are localized within areas of the phase space visited by
predominantly chaotic or regular rays, respectively. Moreover,
Fig. 3c presents a Husimi distribution localized outside the five
sub-islands but in the vicinity of their borders. We conjecture
that this is a finite wavelength analog of the stickiness to the
borders of islands-surround-islands, i.e. the presence of such
parts of the chaotic ray trajectory that correspond to the almost
regular rotation of rays around the islands \cite{ZEN97}. This
occurs when, after wandering in the phase space, the trajectory
approaches sub-islands and \textquotedblleft
sticks\textquotedblright\ to their borders for a fairly long time.
$%
\begin{tabular}
[c]{l}%
\raisebox{-0cm}{\includegraphics[ height=7.7929cm,
width=5.cm,angle=-90
]{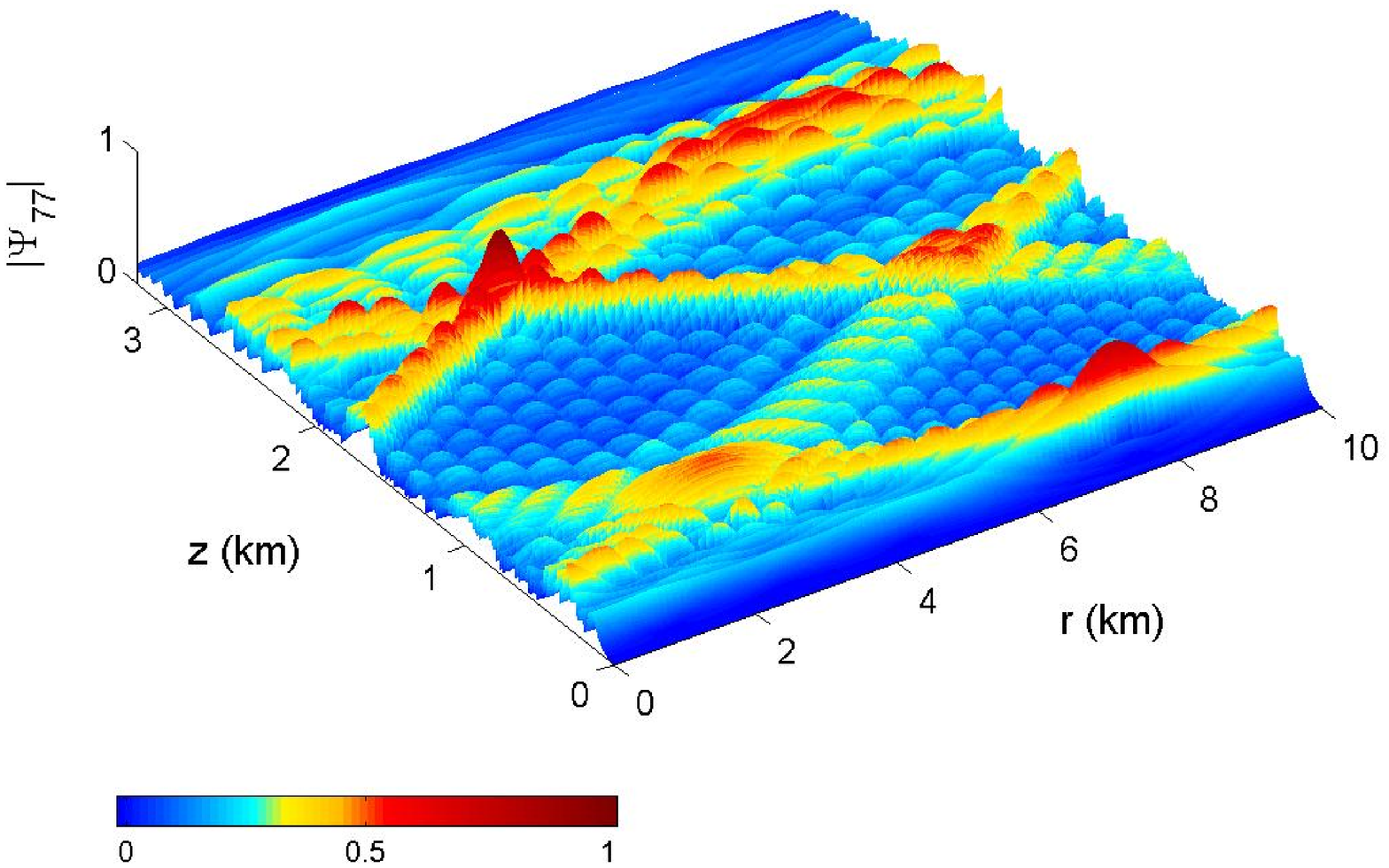}}\\
\raisebox{-0cm}{\parbox[b]{6.761cm}{
\includegraphics[
height=6.761cm, width=5.cm,angle=-90 ]{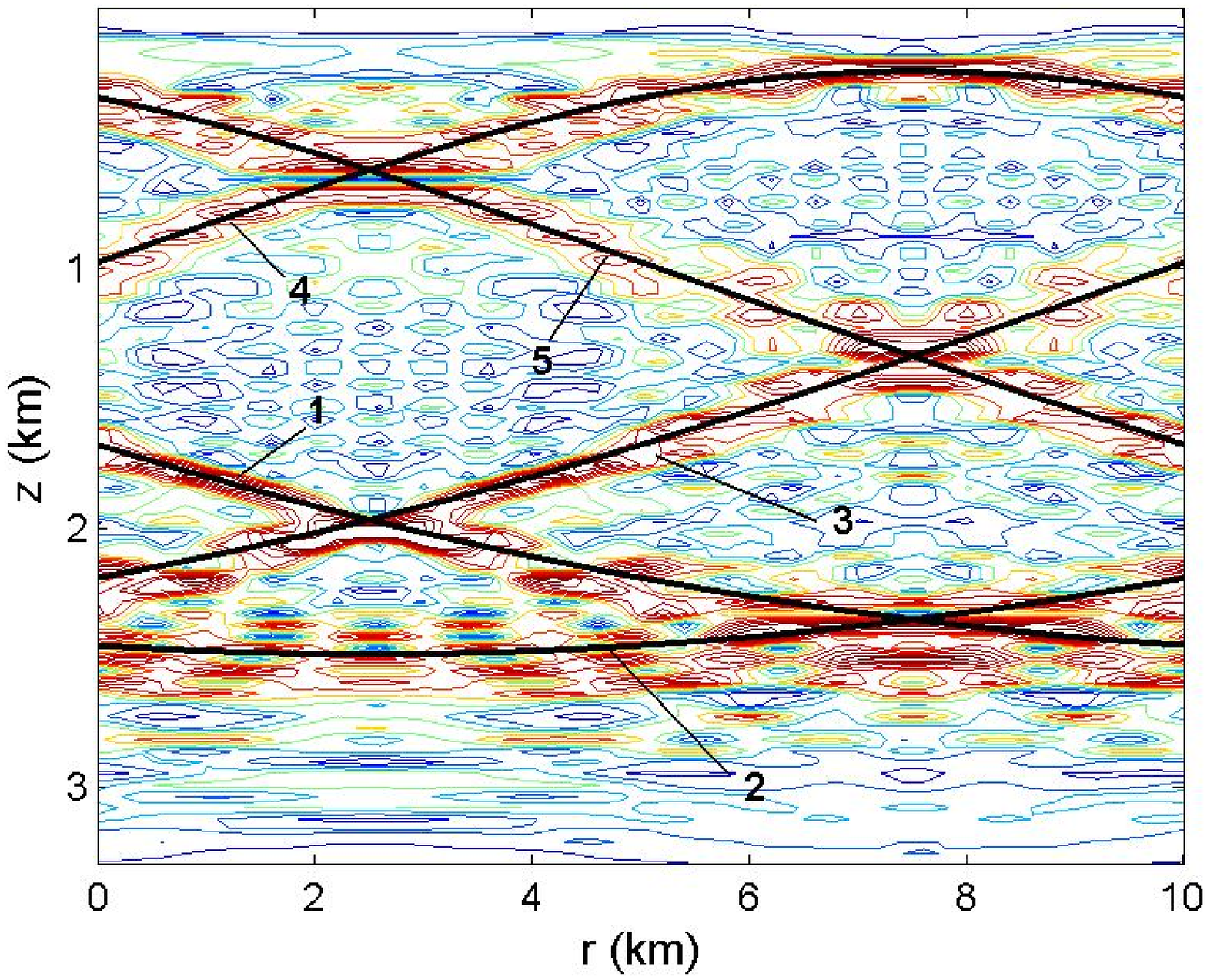} \\ FIG. 5:
Function $| \Psi_{77}(r,z)|$ at a period of its range-dependence.
Five segments of the unstable ray from the upper panel of Fig. 4
are superimposed on the contour plot at the bottom. }}
\end{tabular}$

Figures 3d and 3e need more discussions. Each of them has five
clearly seen maxima. In Fig. 3d the maxima are located within the
chaotic sea, near the destroyed saddles, while in Fig. 3e they are
located near the elliptic points of the sub-islands. We argue that
these maxima are associated with trajectories of two periodic rays
graphed in Fig. 4. Both rays have periods equal to $5L=50$ km,
while there are no other rays with the period of 50 km. The ray
shown in the upper panel of Fig. 4 belongs to the chaotic sea. The
difference $\left\vert \delta z\right\vert $ between its
coordinate and that of a ray with infinitesimally close starting
parameters, on the average, grows as $\exp(\lambda r)$ with the
Lyapunov exponent $\lambda=1/60$ km$^{-1} $. In contrast, the ray
in the lower panel has a zero Lyapunov exponent, i.e. it is
regular. Both rays have been divided into five 10 km segments. The
segments connect the consecutive waveguide cross-sections at range
points $r=nL$ chosen for the stroboscopic observation of both the
ray structure and the Floquet modes (at these range they are
presented by $\psi_{m}(z)$). Small black circles within red and
yellow spots (corresponding to local maxima of Husimi function) in
Fig. 3d indicate positions (in the phase plane) of starting points
of five segments shown in the upper panel of Fig. 4. In agreement
with our expectation these points lie at centers of five maxima of
the Husimi function. The number of each segment is assigned to a
corresponding circle and is indicated next to it in the figure.
Similarly, the five maxima in Fig. 3e are associated with starting
points of segments of the stable ray shown in the lower panel of
Fig. 4. It is natural that the black circles depicting positions
of these points are located in the centers of five stable
sub-islands.
$%
\begin{tabular}
[c]{l}%
\raisebox{-0cm}{\includegraphics[ height=7.7929cm,
width=5.cm,angle=-90.
]{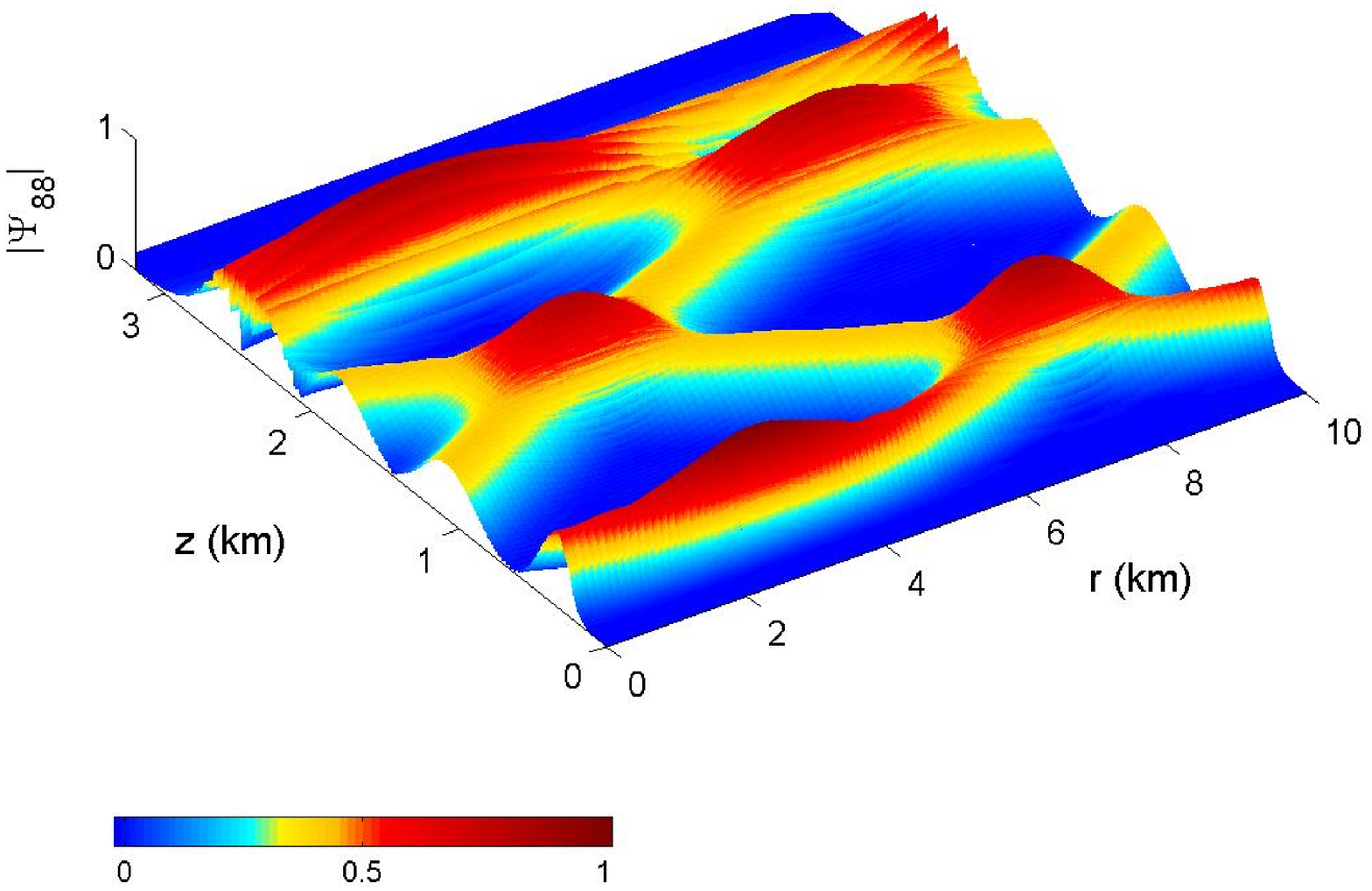}}\\
\raisebox{-0cm}{\parbox[b]{6.761cm}{
\includegraphics[
height=6.761cm, width=5.cm,angle=-90 ]{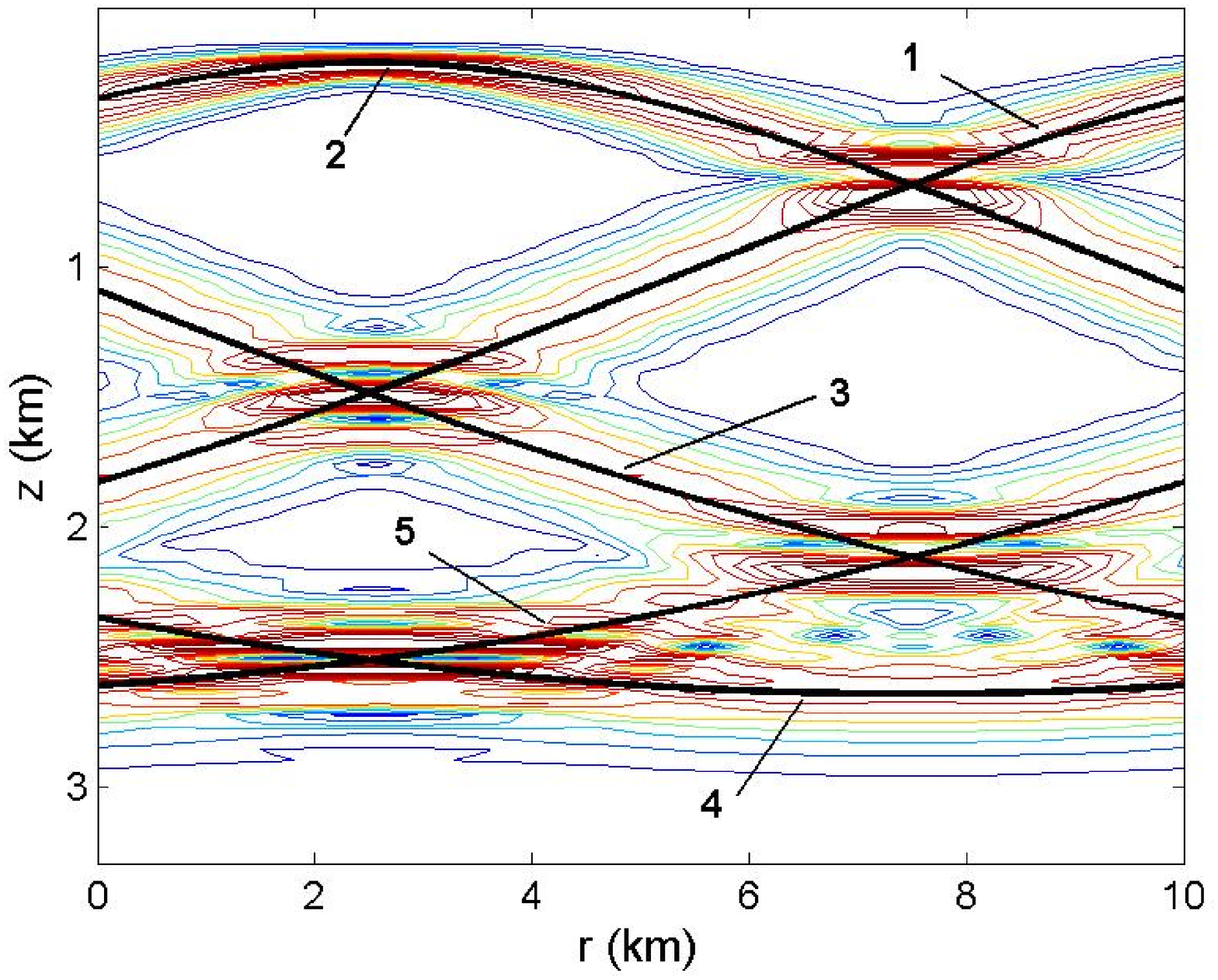}\\ FIG. 6: The
same as in Fig. 5 but for the Floquet mode $\Psi_{88}$. Five
segments superimposed on the contour plot correspond to the stable
ray from the lower panel of Fig. 4. }}
\end{tabular}$

The above results suggest that the two periodic rays shown in Fig.
4 determine essentially the structure of the Floquet modes with
$m=$77 and 88. This conjecture is confirmed in Figs. 5 and 6\ \
where we present distributions of intensities of these modes
within a range interval $(0,L)$, i.e. over a spatial period of all
the Floquet modes. Both functions $\left\vert \Psi
_{77}(r,z)\right\vert $ and $\left\vert \Psi_{88}(r,z)\right\vert
$ , as well as $\left\vert \Psi_{66}(r,z)\right\vert $ shown in
Fig. 7 are scaled to have absolute maxima equal to one. In Fig. 5
we see two representations of function $\left\vert
\Psi_{77}(r,z)\right\vert $. In the lower panel five segments of
the unstable ray from the upper panel of Fig. 4 are superimposed
on the contour plot representing isolines of the intensity
distribution. It is clearly seen that the intensity is high at the
segments of the unstable periodic ray and low elsewhere. So, the
periodic unstable ray path scars the Floquet eigenfunction. This
is an analog to the scarring phenomenon observed in different
autonomous quantum systems \cite{Heller84,BKH2001} (see also in
\cite{Stock99}). In Fig. 6 it is seen that the $88$-th Floquet
mode localized within the five sub-islands also has relatively
large amplitude in the vicinity of segments of the stable ray
shown in the lower panel of Fig. 4. Moreover, practically all
functions $\left\vert \Psi_{m}(r,z)\right\vert $ with $m>50$
display some pattern which may be associated with periodic rays
similar to those shown in Figs. 5 and 6. We illustrate this
statement in Fig. 7 for the Floquet mode with $m=66$.

Figures 5 and 6 suggest that the two Floquet modes may be
considered as a superposition of five 10 km pieces of a wave beams
propagating along the periodic unstable and stable rays,
respectively. We conjecture that scars observed in intensity
distributions of other modes, like that shown in Fig. 7, also may
be associated with segments of some periodic rays. However, this
issue needs a more detailed investigation.
$
\begin{tabular}
[c]{l}%
\raisebox{-0cm}{\includegraphics[ height=7.7929cm,
width=5.cm,angle=-90
]{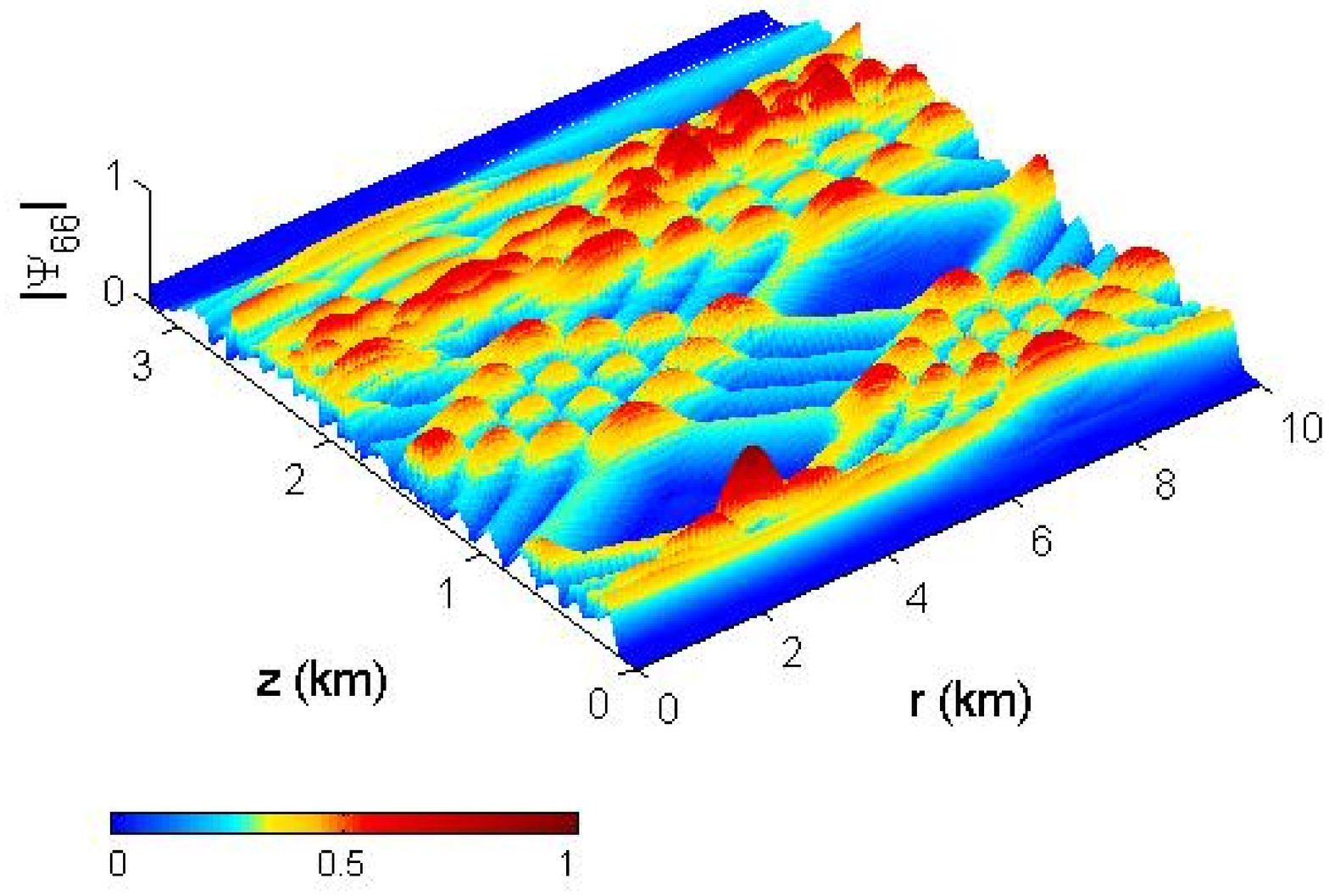}}\\
\raisebox{-0cm}{\parbox[b]{6.761cm}{
\includegraphics[
height=6.761cm, width=5.cm,angle=-90 ]{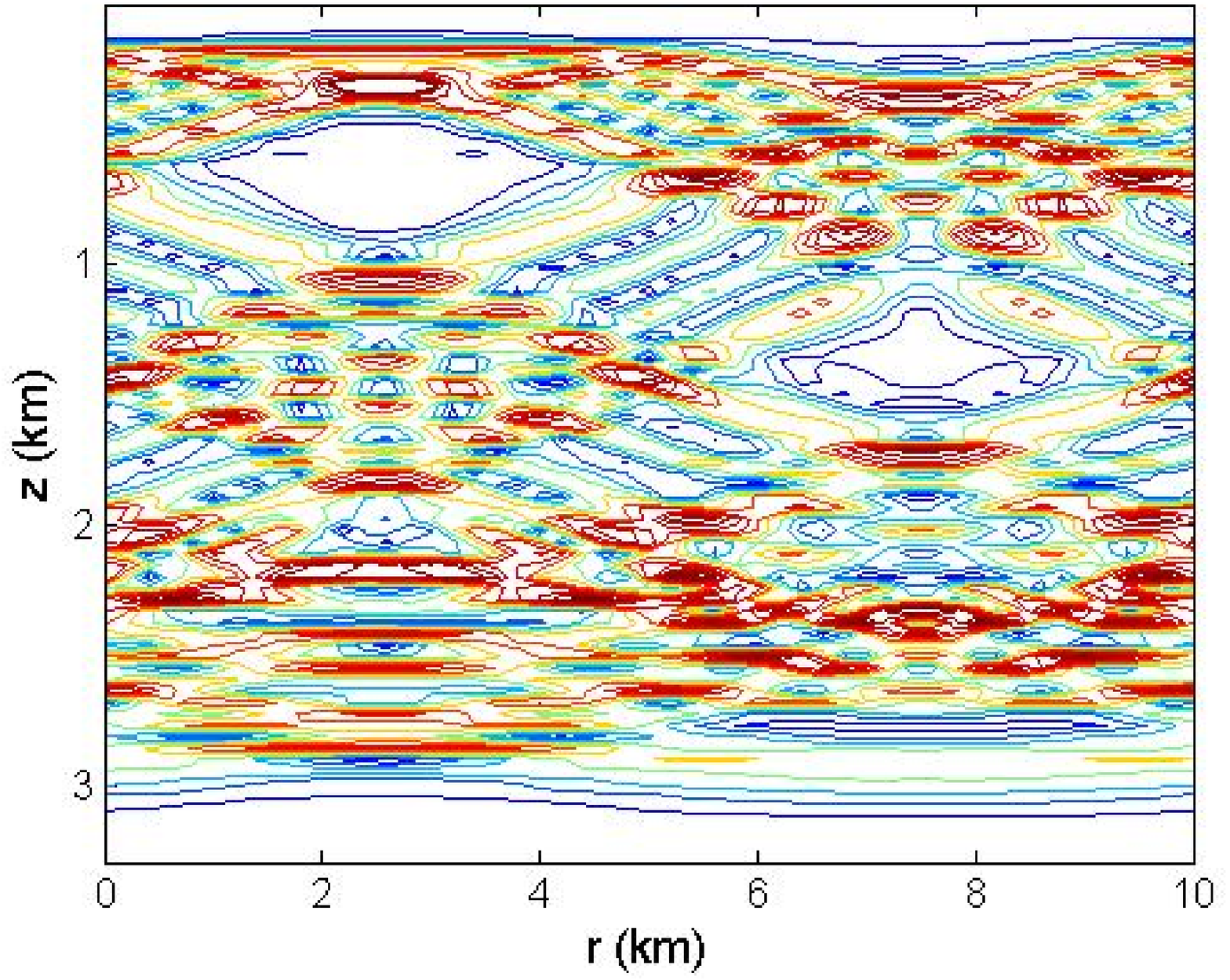}\\ FIG. 7: The
same as in Fig. 5 but for the Floquet mode $\Psi_{66}$ and without
ray segments superimposed on the contour plot. }}
\end{tabular}$

The obtained results demonstrate that the scarring effect may be
observed not only in autonomous quantum systems but also in driven
ones described by the same equations as the waveguides with
periodic range-dependencies. We presume that periodic rays
determine the structures of the Floquet modes in approximately the
same way as periodic orbits determine structures of the wave
functions. The obtained results demonstrate that unstable chaotic
rays can be considered as a reference direction for the
corresponding wave packet propagation. This effect may be used for
the transfer of information and communication.

This work was supported by the U.S. Navy Grants N00014-97-1-0426, and
N00014-02-1-0056, and by the Russian Foundation for Basic Research under Grant
No. 03-02-17246.

\end{document}